\documentclass[12pt]{article}
\usepackage{moriond,psfig}

\def\Journal#1#2#3#4{{#1} {\bf #2}, #3 (#4)}

\def\APJ{\em ApJ}
\def\APJS{\em ApJS}
\def\AA{\em A\&A}

\newcommand\sun{\odot}%
\newcommand\arcdeg{\mbox{$^\circ$}}%
\newcommand\arcmin{\mbox{$^\prime$}}%
\newcommand\arcsec{\mbox{$^{\prime\prime}$}}%
\newcommand\fs{\mbox{$.\!\!^{\mathrm s}$}}%
\newcommand\farcs{\mbox{$.\!\!^{\prime\prime}$}}%
\def\psr{PSR J0218$+$4232\ }

\begin{document}
\vspace*{4cm}
\title{CHANDRA AND RXTE STUDIES OF X-RAY/$\gamma$-RAY MILLISECOND
PULSAR PSR J0218$+$4232: A CLUE TO UNIDENTIFIED $\gamma$-RAY SOURCES?}

\author{L. KUIPER, W. HERMSEN}

\address{SRON-National Institute for Space Research, Sorbonnelaan 2,\\
 3584 CA, Utrecht, The Netherlands}

\maketitle\abstracts{Millisecond pulsar PSR J0218+4232 shows remarkable 
high-energy properties: very hard pulsed X-ray emission up to $\sim 10$ keV and
a likely detection of high-energy $\gamma$-rays ($> 100$ MeV) with a soft spectrum. 
The relative phasing of the X-ray and $\gamma$-ray profiles, however, was 
unknown. A recently performed Chandra (0.08-10 keV) observation of \psr
settled the phasing down to ~0.2 millisecond, and shows that the X-ray pulses 
are aligned with the $\gamma$-ray pulses providing supporting evidence for 
our first detection of high-energy gamma-rays from this source. The preliminary
results from a recent RXTE (2-250 keV) observation (27-12-2001 -- 7-01-2002) 
show significant pulsed emission, in a complex profile, up to $\sim$ 20 keV. 
The composite high-energy spectrum of this millisecond pulsar is similar to the 
canonical spectrum of Unidentified Gamma-ray Sources (UGS), making fast ($<$ 3-4 ms) 
millisecond pulsars with low characteristic ages ($< 10^8$ years) good candidates 
for an UGS association and challenging targets for INTEGRAL observations.
}

\section{Introduction}

\psr is a 2.3 ms pulsar in a two day orbit around a low mass ($\sim 0.2$ M$_{\sun}$) 
white dwarf companion \cite{nav}. Pulsed X-ray emission with a Crab-like double pulse 
profile has been reported from ROSAT 0.1-2.4 keV data \cite{kuip1} and BeppoSAX MECS 
1.6-10 keV data \cite{min}. The pulsed spectrum as measured by the MECS appeared to be 
remarkably hard with a power-law photon index $0.61 \pm 0.32$, harder than measured 
for any other radio pulsar.
Furthermore, Kuiper et al. \cite{kuip2} report the detection with EGRET of pulsed high-energy 
(0.1-1 GeV) $\gamma$-ray emission from this millisecond pulsar.
They also showed that the two $\gamma$-ray pulses appeared to be aligned in 
absolute phase with two of the three radio pulses detected at 610 MHz.
Unfortunately, the timing accuracies of the ROSAT and BeppoSAX observations were
insufficient to construct X-ray profiles in absolute phase. 

\psr is also remarkable in that it is the only Crab-like ms pulsar with a large
DC (unpulsed) fraction of $63 \pm 13\%$ in the ROSAT band below 2.4 keV
\cite{kuip1}, as well as a large DC fraction of $\sim 50\%$ in radio, systematically
over the range 100-1400 MHz \cite{nav}. 
The DC components as measured in the ROSAT and radio observations could be explained by 
emission from a compact nebula with diameter $\sim 14\arcsec$, but in both cases the
indications were at the limit of the imaging capabilities.
Assuming that the radio DC component is compact, combined with the measured very broad 
and structured radio pulse profile, Navarro et al. \cite{nav} suggested that the magnetic 
field of \psr is almost aligned with the rotation axis, the observer viewing the system 
under a small angle with respect to the rotation axis.
Stairs et al. \cite{stairs} determined the magnetic inclination angle analyzing radio polarization 
profiles. Their rotating vector model fits indicate that the magnetic
inclination angle is indeed consistent with $0\arcdeg$ ($8\arcdeg\pm11\arcdeg$). 
Unfortunately, in their fits the line-of-sight inclination is unconstrained. If the DC 
component in X-rays is also compact, for the suggested geometry of a nearly aligned rotator 
and a small viewing angle, it can originate in the pulsar magnetosphere as well as from a 
heated polar cap of the neutron star. 

The objectives of our Chandra and RXTE observations were: 1) To establish the spatial extent of 
the X-ray DC component, compact or extended (Chandra); 2) To construct an X-ray pulse profile which can be 
compared in absolute phase with radio profiles and the EGRET high-energy (0.1-1 GeV)
$\gamma$-ray profile (Chandra and RXTE); and 3) To investigate the characteristics of the pulsed 
emission for energies beyond 10 keV (RXTE).
 

\begin{figure*}[t]
\hbox{\hspace{-0.2cm}{\psfig{file=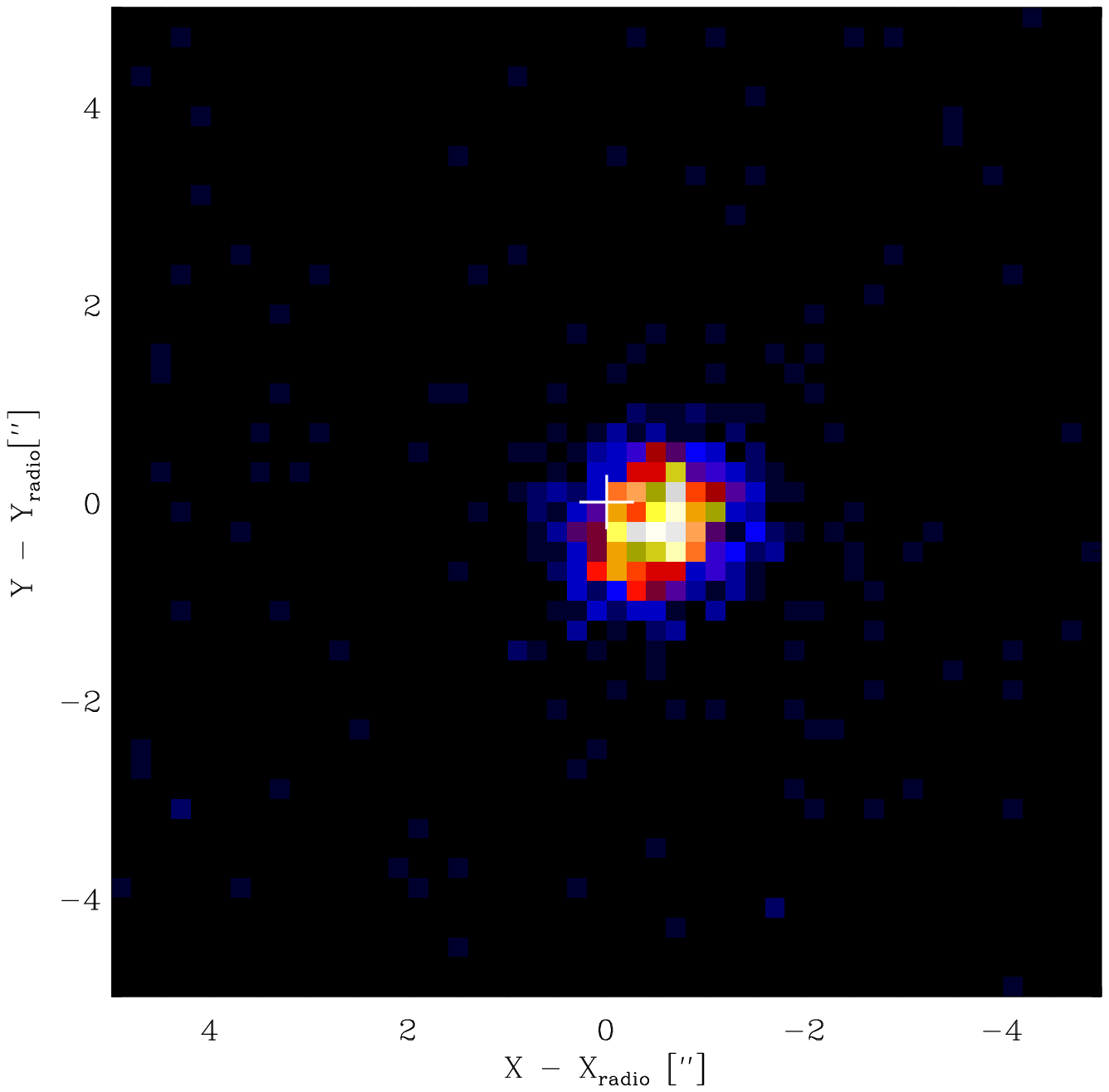,width=8.0cm,height=8.0cm}} 
      \hspace{0.2cm}{\psfig{file=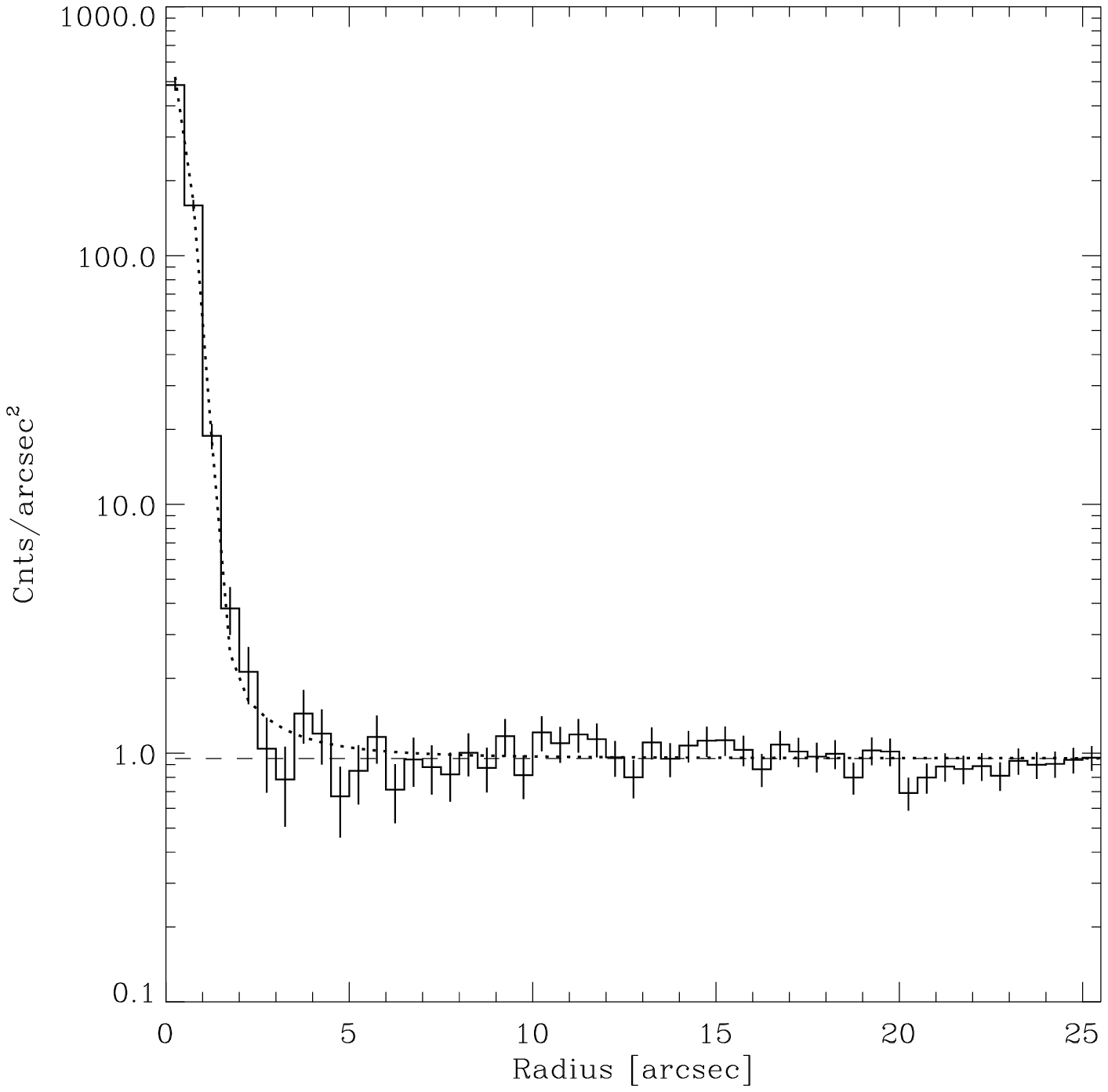,width=8.0cm,height=7.5cm}}}
\caption{(left) Chandra 0.08-10 keV HRC-I image of a $10\arcsec\times 10\arcsec$ 
         region centered on the radio pulsar position of PSR J0218$+$4232. The radio position is marked
         with a `+' sign. The angular distance between the radio pulsar position and the X-ray 
         centroid is $\sim 0\farcs 6$, consistent with the Chandra localization accuracy. 
         (right) Radial distribution of HRC-I events using the optimum X-ray position as
         centre. Superposed as dotted line is the radial profile of the PSF. The dashed line indicates the background 
         level derived from counts in the range 10 -- 25 arcsec from the centre. We have {\bf no} indications
         for extended emission at $\sim 1\arcsec$ scales.\hfill \label{hrcispatial} }
\end{figure*}


\section{Observations}

\psr was observed with the HRC-I and HRC-S instruments on the Chandra
X-ray observatory (CXO) during two observations of $\sim 75$ ks duration each. 
The HRCs are multi-channel plate (MCP) detectors sensitive to X-rays in the 0.08-10 keV energy range 
with no spectral information.
The first observation with the HRC-I camera took place on 1999 December 22 for an effective
exposure of 74.11 ks. Unfortunately, this observation suffered from a non-recoverable timing
(``wiring'') problem assigning incorrectly the event trigger time to that of the next
trigger. The timing accuracy in this case is much worse than the $16\mu$s intrinsic timing
resolution. This degradation prevents the construction of high-resolution pulse profiles in case of
millisecond pulsars.  
In a re-observation of \psr on 2000 October 5 for an effective exposure of 73.21 ks with the HRC-S 
in imaging mode the intrinsic timing accuracy of $16\mu$s could be recovered.

RXTE observed \psr from 27-12-2001 to 7-01-2002 for approximately 200 ks. Data from the PCA (2-60 keV) aboard RXTE
were obtained in the Good Xenon mode, time tagging each trigger with a $0.9 \mu$s time resolution. The two detector
clusters of HEXTE (10-250 keV) operated in staring mode, and the HEXTE science mode was {\bf E\_8us\_256\_DX1F}
allowing spectroscopic studies with 256 channels and a time tag resolution of $7.6 \mu$s.
 
\section{Imaging analysis with Chandra HRC-I}

\psr has clearly been detected near the centre of the $30\arcmin\times 30\arcmin$ field of view in 
the 74.11 ks HRC-I observation. A zoom-in at the pulsar location (Fig. \ref{hrcispatial}
left) shows that the X-ray centroid has a $\sim 0\farcs 6$ offset from the radio-pulsar position of $2^{\mathrm
h}18^{\mathrm m}6\fs 351$, $42\arcdeg 32\arcmin 17\farcs 45$ (epoch J2000), well within the celestial 
localization accuracy determination requirement of $1\arcsec$. The radial profile is compatible with the PSF of the HRMA/HRC-I
combination (95\% of the source counts are within $2\arcsec$ from the X-ray centroid). Fig. \ref{hrcispatial}
(right) shows the best model profile (dotted line) superposed on the measured radial profile using the optimum X-ray
centroid position as centre. Thus, we have no evidence for extended emission near \psr at $\sim 1\arcsec$ scales
(diameter), rejecting the indication for a compact nebula found in our analysis of ROSAT HRI data
\cite{kuip1}. 


\begin{figure*}[h]
  \hbox{\hspace{0.0cm} 
        \psfig{figure=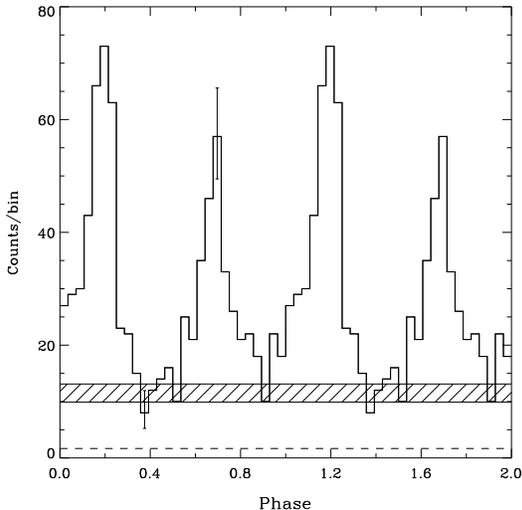,width=7.5cm,height=7.5cm}
        \hspace{0.2cm}{\parbox{78mm}{\vspace{-7cm}
        \caption{Pulse-profile of \psr as measured by the HRC-S in imaging
           mode. Two cycles are shown for clarity. Typical error bars 
           are shown at phases 0.35 and 0.7.
           The DC-level ($\pm 1\sigma$) is indicated by the hatched area, and the
           background level from the imaging analysis by the dashed line.\hfill
           \label{hrcspulseprof}}}}}
\end{figure*}


\section{Timing analysis with Chandra HRC-S and RXTE PCA}

The first step in the timing analysis of the Chandra HRC-S data is to correct the assigned event
times by back-shifting, recovering the intrinsic relative time resolution of $16\mu$s. Next, the event
extraction radius was set to the optimum radius of $1\farcs 5$. The final step is the determination of 
the arrival times at the solar system barycentre using the orbital information of Chandra and the position of 
PSR J0218$+$4232. Folding the barycentered arrival times of the selected events with the spin and binary
parameters from an updated ephemeris \cite{kuip3} for \psr revealed the well-known doubled peaked profile 
at high statistics (Fig.\ref{hrcspulseprof}).

The deviation from a flat distribution is $15.2\sigma$ according to a $Z^{2}_{6}$ - test and
the peak separation is $0.475\pm0.015$ consistent with previous estimates \cite{kuip1,min}.
In Fig.\ref{hrcspulseprof} an estimate for the unpulsed (DC) level ($\pm 1\sigma$) is indicated 
using a bootstrap method outlined by Swanepoel et al. \cite{swan} Furthermore, the background level
determined in an imaging analysis is also shown. The pointlike DC (unpulsed)
component, significantly detected with ROSAT (0.1-2.4 keV; $4.8 \sigma$) by Kuiper et al.\cite{kuip1} 
and weakly with BeppoSAX (1.6-4.0 keV; $\sim 2 \sigma$) by Mineo et al.\cite{min} is clearly seen 
in this 0.08-10 keV profile. The pulsed fraction $\cal{F}_P$ defined as ${\cal{N}_P}/({\cal{N}_P+\cal{N}_{DC}})$, 
where $\cal{N}_P$ specifies the number of pulsed source counts and $\cal{N}_{DC}$ the number of DC source counts, 
turns out to be ${\cal{F}_P} = 0.64\pm0.06$ as measured over the entire 0.08-10 keV energy range.
\begin{figure*}[h]
  \hbox{\hspace{0.0cm} 
        \psfig{figure=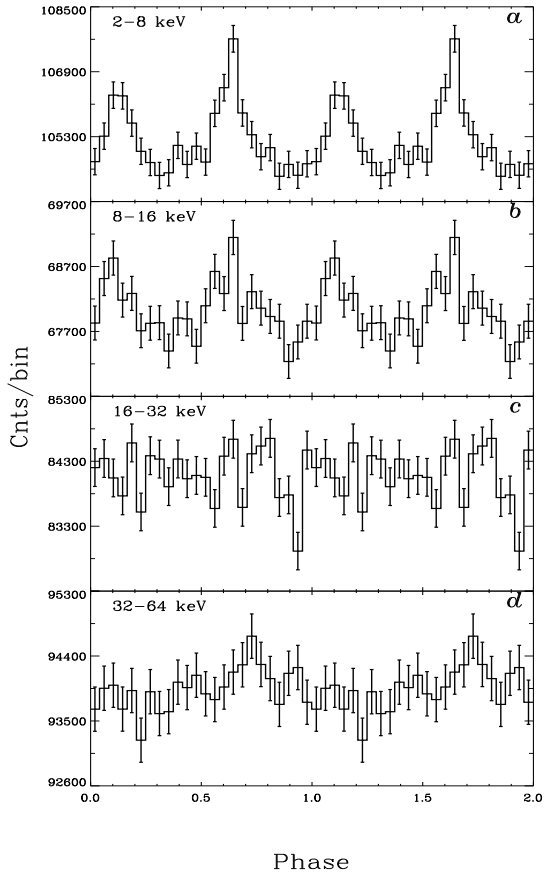,width=7.5cm,height=12.75cm,bbllx=130pt,bblly=120pt,bburx=420pt,bbury=680pt,clip=}
        \hspace{0.4cm}{\parbox{76mm}{\vspace{-11cm}
        \caption{RXTE PCA pulse profiles of \psr in 4 different energy window obtained from a $\sim$ 
                 200 ks observation. Significant pulsed emission has been detected up to at least $\sim$
                 16 keV.\hfill
                 \label{pcapulseprof}}}}}
\end{figure*}
\noindent The barycentered PCA trigger times (no screening of PCA data has yet been applied) are phase folded according 
to the same updated ephemeris as used before. The resulting profiles in 4 differential energy windows are shown
in Fig.\ref{pcapulseprof}. The significances for deviations from flat distributions according 
to a $Z_4^2$ test are: $10.9\sigma, 5.5\sigma, 1.4\sigma$ and $1.5\sigma$ for the 2-8, 8-16, 16-32 and 32-64 keV energy windows, 
respectively. Thus, for the first time we measured a significant pulsed signal for energies above 10 keV (see 
Fig.\ref{pcapulseprof}{\it b}). The profiles shown in Fig. \ref{pcapulseprof}{\it c,d} do {\bf not} show significant
timing signals.

It is now also possible to cross-calibrate the Chandra and RXTE profiles in the overlapping energy range. Taking into account
the internal delays (PCA 16$\mu$s; Chandra HRC-S 19.5$\mu$s) we found that the Chandra clock runs $105\pm 17\mu$s ahead the
RXTE clock. This significantly improves upon a previous estimate based on a Chandra-RXTE correlation study of the much slower 
rotating Crab pulsar \cite{ten}. In the latter study it was found that the Chandra clock runs behind the RXTE clock by
$200\pm 100\mu$s.

\begin{figure*}
  \hbox{\hspace{0.0cm} 
        \psfig{figure=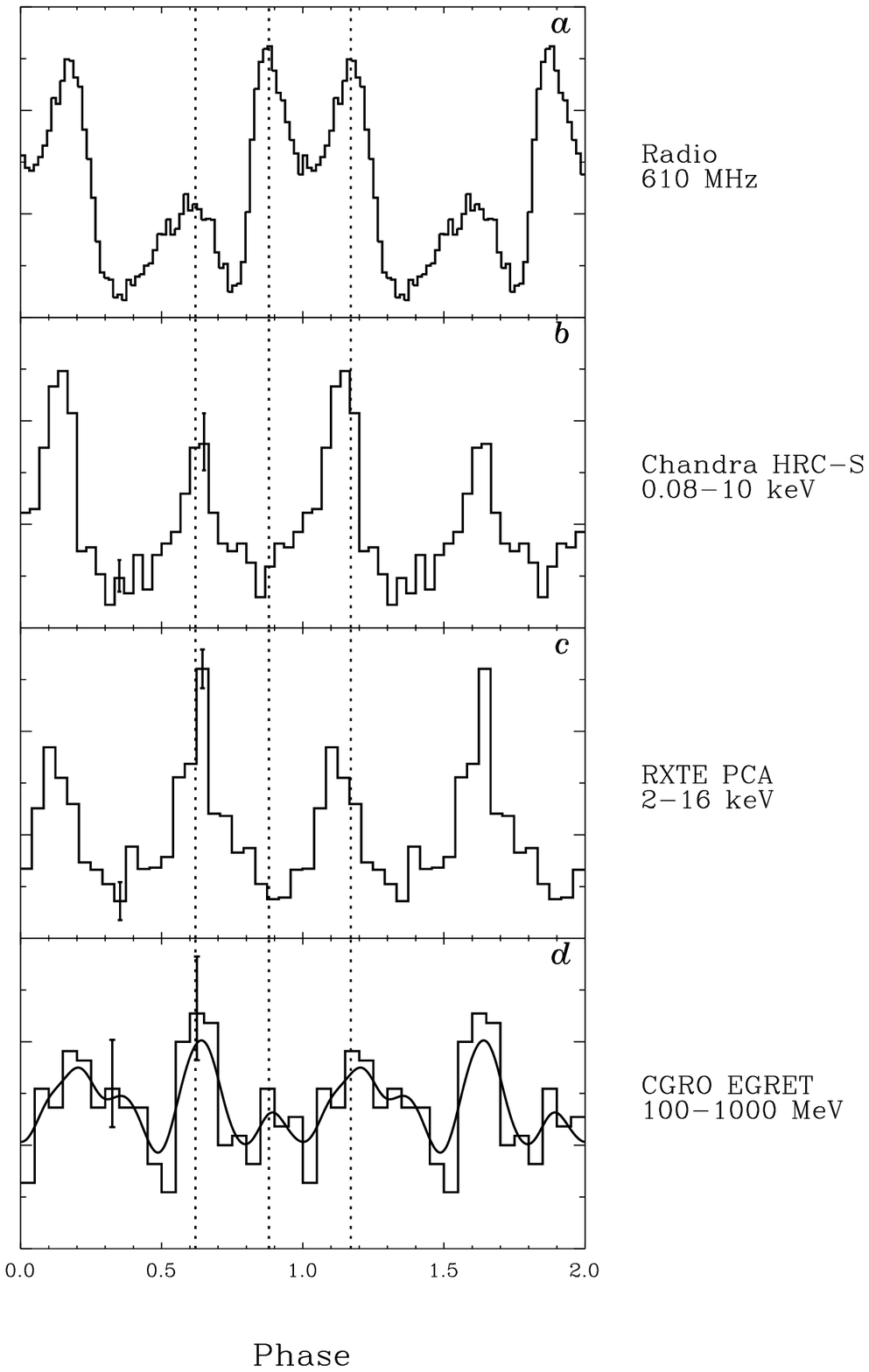,width=7.cm,height=12.75cm,bbllx=165pt,bblly=120pt,bburx=420pt,bbury=680pt,clip=}
        \hspace{0.4cm}{\parbox{78mm}{\vspace{-13cm}
        \caption{Multi-wavelength pulse profiles of \psr in absolute phase. {\it a}) Radio profile at 610 MHz; {\it b}) Chandra
        HRC-S 0.08-10 keV pulse profile (mean weighted energy at $\sim 1.7$ keV); {\it c}) RXTE PCA 2-16 keV profile (mean 
        weighted energy at $\sim 7.6$ keV); {\it d}) 0.1-1 GeV CGRO EGRET pulse profile with superposed the kernel density 
        estimator. The vertical dashed lines indicate the location of the three radio pulses. The 2 non-thermal X-ray/
        $\gamma$-ray pulses are aligned with 2 of the 3 radio pulses.\hfill
        \label{mwvpulseprof}}}}}
\end{figure*}

\begin{figure*}[h]
  \hbox{\hspace{0.0cm} 
        \psfig{figure=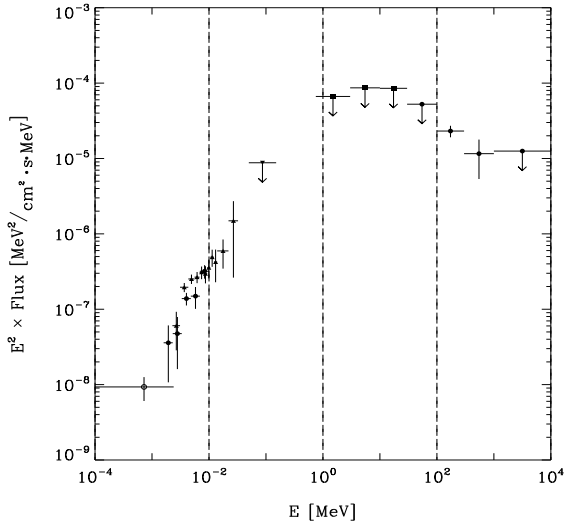,width=7.5cm,height=7.0cm}
        \hspace{0.2cm}{\parbox{78mm}{\vspace{-7cm}
        \caption{High-energy total pulsed spectrum of \psr from soft X-rays up to high-energy $\gamma$-rays.
                 The newly derived RXTE fluxes (filled upwards pointing triangles) are combined with spectral
                 information from the ROSAT HRI (0.1-2.4 keV, open circle), BeppoSAX MECS (2-10 keV, filled circles),
                 CGRO OSSE (50-150 keV, filled downwards pointing triangle), CGRO COMPTEL (0.75-30 MeV, filled squares) and
                 CGRO EGRET (30 - 10000 MeV, filled circles).\hfill
           \label{hespc}}}}}
\end{figure*}

Fig.\ref{mwvpulseprof} shows {\bf for the first time} in absolute phase the comparison of the \psr Chandra HRC-S (back shift 
of $0.045$ in phase applied) and RXTE PCA X-ray profiles with the radio and $\gamma$-ray profiles. The non-thermal X-ray pulses appear to be
aligned within the timing uncertainties with two of the three radio pulses and with the two $\gamma$-ray pulses, strengthening the 
credibility of the earlier reported first detection of pulsed high-energy $\gamma$-ray emission from a (this) millisecond pulsar \cite{kuip2}.
In Kuiper et al. \cite{kuip3} we show that the $\gamma$-ray detection significance increases to $4.9\sigma$.
Note the morphology change of the X-ray profile already known from the BeppoSAX study\cite{min}.
We also produced a (preliminary) RXTE PCA total pulsed spectrum using template fitting (the best statistics 2-16 keV profile
was used as template for all differential energy windows). The results are combined with the spectral findings from previous high-energy studies in Fig. \ref{hespc}. The new RXTE data support a continuation of the hard (photon Pl-index of about -1)
total pulsed spectrum beyond 10 keV. The spectrum, hard at X-rays and soft at high-energy $\gamma$-rays, is similar to that
of Unidentified Gamma-Ray Sources (UGS)\cite{grenier}. Therefore fast millisecond pulsars, particularly those with low
characteristic ages, seem to be promising candidates for an UGS association.


\section{Conclusions}

From the recently performed deep Chandra and RXTE observations of \psr we learned that:\\
$\bullet$ \hspace{0.2cm} There is no evidence for extended emission at X-ray energies at $\sim 1\arcsec$ scales.\\
$\bullet$ \hspace{0.2cm} The non-thermal X-ray pulses are aligned with 2 of the 3 radio pulses and with 
the two high-energy $\gamma$-ray pulses strengthening the credibility of the first detection of pulsed high-energy 
$\gamma$-ray emission from a (this) millisecond pulsar.\\
$\bullet$ \hspace{0.2cm} The total pulsed X-ray spectrum continues up to $\sim 20$ keV into the soft $\gamma$-ray 
domain.\\
\noindent Future observations of this intriguing millisecond pulsar at soft $\gamma$-rays by INTEGRAL and at high-energy $\gamma$-rays by
AGILE and GLAST are very important to determine its spectral characteristics in the $\gamma$-ray regime in more detail and possibly to 
shed light on the origin of UGSs of which \psr could be a canonical prototype.

\end{document}